\documentclass{iau} 
\usepackage{graphicx}
\usepackage{amsmath, amsfonts, amssymb, cancel, gensymb, graphicx, float, parskip, subcaption}

\title[Variability and Binarity at the GC] %% give here short title %%
{Constraining the Variability and Binary Fraction of Galactic Center Young Stars}

\author[Gautam et al.]   %% give here short author list %%
{Abhimat K. Gautam$^1$, Tuan Do$^1$, Andrea M. Ghez$^1$, Jessica R. Lu$^2$, Mark R. Morris$^1$, Shoko Sakai$^1$, Gunther Witzel$^1$, Breann~N.~Sitarski$^1$, and Samantha Chappell$^1$}

\affiliation{$^1$Dept. of Physics and Astronomy, University of California, Los Angeles \\Los Angeles, CA, 90024, USA\\ email: {\tt abhimat@astro.ucla.edu} \\[\affilskip]
$^2$Dept. of Astronomy, University of California, Berkeley\\Berkeley, CA, 94720, USA}

\pubyear{2016}
\volume{322}  %% insert here IAU Symposium No.
\jname{The Multi-Messenger Astrophysics of the Galactic Centre}
\editors{R. Crocker, S. Longmore, \& G. Bicknell, eds.}
\begin{document}

\maketitle

\begin{abstract}
We present constraints on the variability and binarity of young stars in the central 10~arcseconds ($\sim 0.4$ pc) of the Milky Way Galactic Center (GC) using Keck Adaptive Optics data over a 12 year baseline. Given our experiment's photometric 
uncertainties, at least $36\%$ of our sample's known early-type stars are variable. We identified eclipsing binary systems by searching for periodic variability. In our sample of 
spectroscopically confirmed and likely early-type stars, we detected the two previously discovered GC eclipsing binary systems. We derived the likely binary fraction of main 
sequence, early-type stars at the GC via Monte Carlo simulations of eclipsing binary systems, and find that it is at least $32\%$ with $90\%$ confidence.
\end{abstract}

% \firstsection % if your document starts with a section,
              % remove some space above using this command.

In this work, we better constrain the Milky Way Galactic Center (GC) binary fraction with much larger samples, improving upon previous studies limited to the brightest sources. Tighter binary fraction constraints are especially useful to understand the formation and dynamics of young stars in the nuclear star cluster near the central supermassive black hole: The fragmentation of an accretion disk around the black hole could have formed the young stars currently observed in the clockwise disk at the GC (\cite[Alexander et al. 2008]{2008MNRAS.389.1655A}). The binary fraction also constrains the binary disruption mechanism (\cite[Hills 1988]{1988Natur.331..687H}) that could explain the origin of the observed young S-stars within $\sim 0.05$ pc of the central supermassive black hole.

We used laser guide-star adaptive optics high-resolution images of the GC obtained at the 10 m. W. M. Keck II telescope with the NIRC2 near-infrared facility imager through the $K'$ ($\lambda_0 = 2.124$ $\mu$m) bandpass. The images were centered near the position of Sgr~A* and span 10 arcseconds ($\sim 0.4$ pc at $R_0 \approx 8.0$ kpc), and were taken over 37 nights between July 2004 and May 2016 (observation setup used here is further detailed by \cite[Ghez et al. 2008]{2008ApJ...689.1044G} and \cite[Yelda et al. 2014]{2014ApJ...783..131Y}).

Variable stars were identified as those displaying photometric fluctuations above that expected from $7\sigma$ noise fluctuations. At least 36\% of the 72 known early-type stars identified in all nights are variable.

\begin{figure}[h]
  \centering
  \begin{subfigure}{.5\textwidth}
    \centering
    \includegraphics[width=\linewidth]{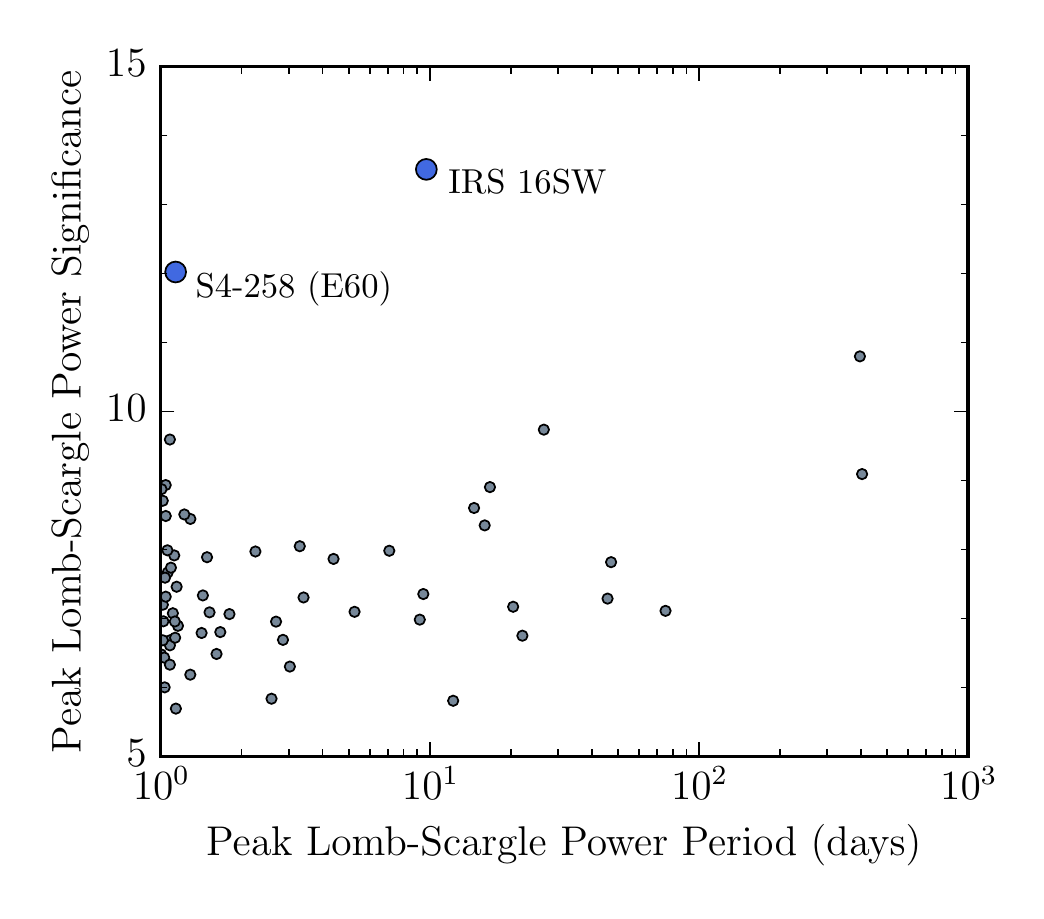}
    \label{fig:LS_sig}
  \end{subfigure}%
  \begin{subfigure}{.5\textwidth}
    \centering
    \includegraphics[width=\linewidth]{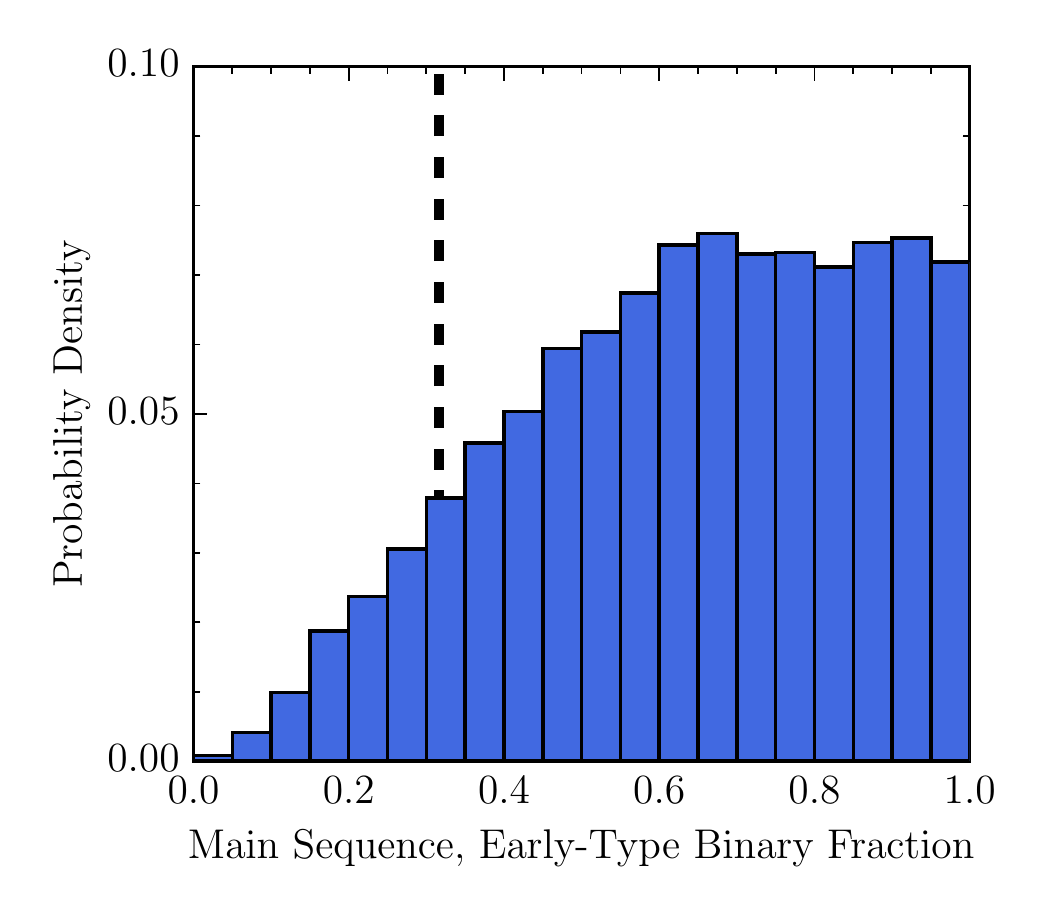}
    \label{fig:Bin_Frac}
  \end{subfigure}%
  \caption{\label{fig:Binaries}\emph{Left:} The two previously discovered eclipsing binary systems in the central $10''$ were identified via a Lomb-Scargle search for periodicity: IRS 16SW (\cite[Ott et al. 1999]{1999ApJ...523..248O}, \cite[Rafelski et al. 2007]{2007ApJ...659.1241R}) and S4-258 (E60: \cite[Pfuhl et al. 2014]{2014ApJ...782..101P}). \emph{Right:} Main sequence, early-type eclipsing binaries were simulated with randomly distributed inclinations. Initial mass function for the primary star followed \cite[Lu et al. 2013]{Lu:2013en}, and mass and separation distributions followed \cite[Sana et al. 2012]{2012Sci...337..444S}. Detectability of trial binary systems was determined by comparing maximum eclipse depths to our photometric uncertainty limits. The probability distribution of the main sequence, early-type binary fraction with 2 binaries detected is shown. We constrain it to be at least $32\%$ with 90\% confidence (dashed line).}
\end{figure}

To identify eclipsing binaries, we conducted a periodic variability search on 124 spectroscopically confirmed or likely early-type stars ($p(\text{Early-Type}) \geq 0.5$, from \cite[Do et al. 2013]{2013ApJ...764..154D}) identified in at least 25 nights. We constrained the binary fraction with a Monte Carlo simulation of eclipsing binaries, detailed in Figure~\ref{fig:Binaries}. We find that the GC main sequence, early-type binary fraction is at least $32\%$ with $90\%$ confidence. This work will be further detailed by Gautam et al., in prep.

\end{document}